 \documentclass[12pt,apj]{emulateapj}

\usepackage{lineno}
\usepackage{natbib}
\usepackage{amsmath}
\usepackage{graphicx}
\usepackage{wasysym}
\usepackage{txfonts}
\usepackage{xspace}
\usepackage{url}
\usepackage{textcomp}
\usepackage{multirow}
\usepackage{lipsum}

\newcommand{\ngc}{NGC\,5907~ULX1\xspace}

\newcommand{\swift}{\textsl{Swift}\xspace}

\newcommand{\chandra}{\textsl{Chandra}\xspace}
\newcommand{\xmm}{\textsl{XMM-Newton}\xspace}

\newcommand{\nustar}{\textsl{NuSTAR}\xspace}

\newcommand{\asec}{\ensuremath{''}\xspace}

\newcommand{\snr}{S/N\xspace}

\newcommand{\msun}{\ensuremath{\text{M}_{\odot}}\xspace}

\renewcommand{\deg}{\ensuremath{^\circ}}
\newcommand{\ergcms}{\ensuremath{\text{erg\,cm}^{-2}\text{s}^{-1}}}

\shorttitle{Spectral changes in \ngc}
\shortauthors{F\"urst et al.}

\begin{document}



\title{Spectral Changes in the Hyperluminous Pulsar in NGC\,5907 as a Function of Super-Orbital Phase}

\author{F.~F\"urst\altaffilmark{1}}
\author{D.~J.~Walton\altaffilmark{2,1}}
\author{D.~Stern\altaffilmark{2}}
\author{M.~Bachetti\altaffilmark{3}}
\author{D.~Barret\altaffilmark{4}}
\author{M.~Brightman\altaffilmark{1}}
\author{F.~A.~Harrison\altaffilmark{1}}
\author{V.~Rana\altaffilmark{1}}

\altaffiltext{1}{Cahill Center for Astronomy and Astrophysics, California Institute of Technology, Pasadena, CA 91125, USA; \email{fuerst@caltech.edu}}
\altaffiltext{2}{Jet Propulsion Laboratory, California Institute of Technology, Pasadena, CA 91109, USA}
\altaffiltext{3}{INAF/Osservatorio Astronomico di Cagliari, via della Scienza 5, I-09047 Selargius (CA), Italy}
\altaffiltext{4}{IRAP/CNRS, 9 Av. colonel Roche, BP 44346, F-31028 Toulouse cedex 4, France and
Universit\'e de Toulouse III Paul Sabatier / OMP, Toulouse, France}

\begin{abstract}

We present  broad-band, multi-epoch X-ray spectroscopy of the pulsating ultra-luminous X-ray source (ULX) in NGC\,5907.  Simultaneous \xmm and \nustar data from 2014 are best described by a multi-color black-body model with a temperature gradient as a function of accretion disk radius significantly flatter than expected for a standard thin accretion disk ($T(r)\propto r^{-p}$, with $p=0.608^{+0.014}_{-0.012}$). Additionally, we detect a hard power-law tail at energies above 10\,keV, which we interpret as being due to Comptonization. We compare this observation to  archival \xmm, \chandra, and \nustar data from 2003, 2012, and 2013, and investigate possible spectral changes as a function of  phase over the 78\,d super-orbital period of this source. We find that observations taken around phases 0.3--0.4 show very similar temperature profiles, even though the observed flux varies significantly, while one observation taken around phase 0 has a significantly steeper profile. 
We discuss these findings in light of the recent discovery that the compact object is a neutron star and show that precession of the accretion disk or the neutron star can self-consistently explain most observed phenomena.

\end{abstract}

\keywords{X-rays: binaries --- pulsars: individual (NGC 5907 ULX1) --- accretion, accretion disks}

\section{Introduction}
\label{sec:intro}

 Only a few sources in our Galaxy are known to be able to sustain
luminosities close to the Eddington luminosity. However, in nearby
galaxies many dozens of off-nuclear systems are known that reach
luminosities greater than $1.4 \times 10^{39}\, {\rm erg}\, {\rm
s}^{-1}$, the Eddington luminosity for a typical black hole binary
with a mass of $M_{\rm BH} = 10\,\msun$; some of these exceed
this luminosity by orders of magnitude  \citep[e.g.,][]{swartz08a, walton11a}.  Since these sources are significantly
separated from the center of mass of their host galaxies, they
cannot be related to 7super-massive black holes.
We refer to these extreme accretors as ultra-luminous
X-ray sources (ULXs).

Due to their high luminosity, it has been speculated that ULXs host intermediate mass black holes \citep[IMBHs, see, e.g.,][]{colbert99a}, which might provide important building blocks to form  the super-massive black holes that power active galactic nuclei \citep{volonteri10a}.
  In this case, and assuming accretion physics is largely mass-invariant, we would expect a power-law like hard X-ray spectrum, 
  with a roll-over at energies $\gg100$\,keV produced by Comptonization in a hot corona.
However, numerous studies  have shown that  most bright ULXs 
show a distinctly different spectral shape \citep[e.g.,][]{stobbart06a, gladstone09a}. In particular, in the \nustar era the high-energy ($>$10\,keV) spectra of bright ULXs have now become routinely observable. Such ULXs show a spectrum  that seems thermal in origin, with a fast turn-over above $\sim10$\,keV. 

The  spectral shape of these ULXs is also distinctly different to the spectra seen in the sub-Eddington accretion regime of  most Galactic binaries and active galaxies.  They are therefore likely stellar remnants accreting above the Eddington rate and typically assumed to black holes.
However, in a  surprising discovery  \citet{bachetti14a} found that M82~X-2 is powered by a neutron star accreting at super-Eddington levels. Recently, two more neutron star powered ULXs were identified through their pulsations: NGC\,7793~P13 \citep{p13, israel16a} and \ngc \citep{israel16b}.

\ngc has been reported to exhibit peak luminosities of up to $6\times10^{40}$\,erg\,s$^{-1}$, assuming a distance of 13.4\,Mpc \citep{sutton13a}. However, the most recent distance estimate by \citet{tully16a} puts NGC\,5907 at an even larger distance of 17.06\,Mpc, which increases the peak luminosity to $\sim10^{41}$\,erg\,s$^{-1}$, and  places \ngc among the extremely rare subset of ULXs referred to as hyper-luminous X-ray sources ($L_{\rm{X}} \geq 10^{41}$\,erg\,s$^{-1}$). Its neutron star nature makes it a completely unique source, which apparently is accreting at about 500 times the Eddington rate \citep{israel16b}.

Using \xmm, \citet{sutton12a, sutton13a} found that the spectrum of \ngc shows a tentative high-energy roll-over at about 5\,keV, in line with expectations from super-Eddington accretion.
\citet[hereafter W15]{walton15a} analyzed broad-band \xmm and \nustar and found a very good fit with a thermal spectrum, confirming the roll-over at high energies.
This spectral shape is similar to many other ULX systems observed by \nustar in recent years \citep[e.g.,][]{bachetti13a, walton13a, walton14a, walton15a, walton15b, rana15a, mukherjee15a}, many which could harbor black holes, due to their lack of pulsations \citep[e.g.,][]{doroshenko15a}. The spectral similarity might indicate that the observed radiation is dominated by the effects of the super-Eddington accretion flow and not by the properties of the compact object. 

W15 obtained two epochs of observations in 2013, but found the source in an ``off-state'' during the first observation, in which it was not detected by \nustar and was only marginally (at best) detected by \xmm. In the second epoch, only 4\,d later, the flux had recovered, having risen by at least 2 orders of magnitude. However, the observed  0.3--20.0\,keV  flux of $\left(7.2 \pm 0.3\right) \times 10^{-13}\,$erg\,s$^{-1}$\,cm$^{-2}$ was still at the low end of typically observed fluxes for this source. 

\begin{figure*}
\begin{center}
\includegraphics[width=0.95\textwidth]{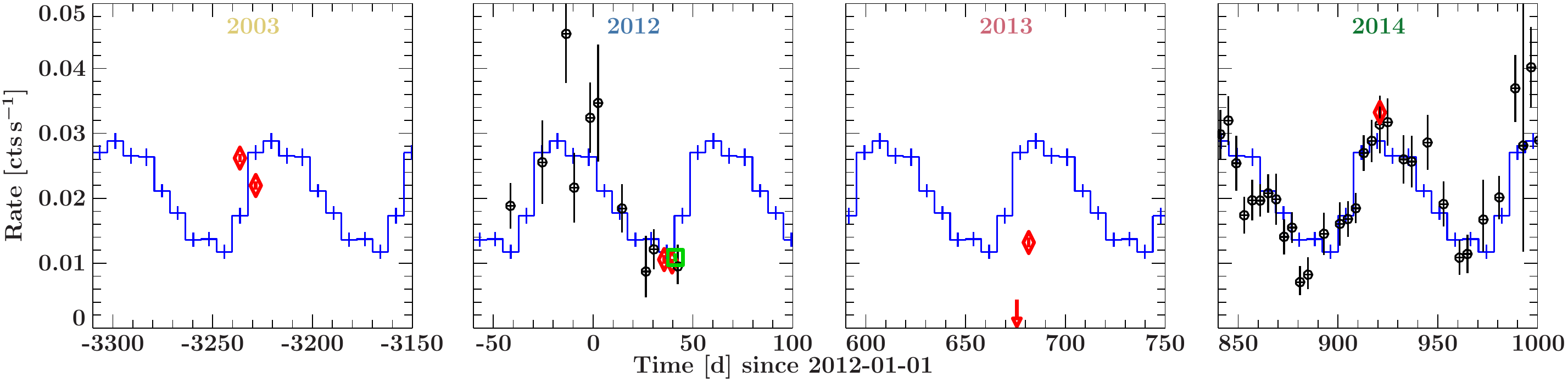}
\caption{Longterm \swift/XRT light-curve of \ngc shown as black circles, together with the four epochs of \xmm observations, shown by the red diamonds and the \chandra observation in 2012 indicated by the green square. The observations in 2013 and 2014 were obtained simultaneously with \nustar. The blue profile is the average \swift/XRT light-curve between 2014--2016 folded on the 78-d period discovered by W16. The average profile was extrapolated back in time assuming a constant period.
The \xmm and \chandra fluxes were scaled to the \swift/XRT rates. }
\label{fig:xrtlc}
\end{center}
\end{figure*}

Following up on the remarkable flux variability observed in \ngc, \citet[hereafter W16]{walton16b} presented results of intense monitoring of the source with \swift/XRT over more than 2 years. While they did not observe another ``off-state'', they found evidence for a stable $\sim$78\,d period. This period is most likely super-orbital, as \citet{israel16b} find evidence in the timing data of \ngc for a $\sim$5\,d orbital period.  Figure~\ref{fig:xrtlc} shows part of the \swift/XRT monitoring light curve with the average profile of the super-orbital period super-imposed. Extrapolating this profile back to the \xmm observations taken in 2003 and 2012, we can see that the 2003 data were taken close to a high-state, while the 2012 data were taken close to the minimum of the profile. The 2013 and 2014 \nustar and \xmm observations all fall close to the expected peak of this cycle, although the two 2013 observations both showed abnormally low fluxes for their phase.

Here we present observations taken simultaneously with \xmm and \nustar during a high state of  the 78\,d period of \ngc in 2014 to obtain a higher quality X-ray spectrum above 10\,keV and search for changes of the spectral parameters as function of luminosity and time. These data are also analyzed by \citet{israel16b}.

The remainder of the paper is structured as follows: in Section~\ref{sec:data} we describe the data reduction and extraction. In Section~\ref{sec:spec} we first analyze the 2014 data in detail and then compare their spectral shape to archival data. We summarize and discuss our results in Section~\ref{sec:disc}.

\section{Data reduction and observation}
\label{sec:data}

Besides the new \nustar and \xmm observations of 2014, we  use three more epochs of observations: \xmm data taken in 2003, \xmm and \chandra data taken in 2012, and \xmm and \nustar data taken in 2013. The former two epochs were first analyzed by \citet{sutton13a} and the 2013 data were presented by W15.
For the remainder of this paper, we identify the epochs by the year they occurred in; see Table~\ref{tab:obslog}. 

\begin{deluxetable}{lcccc}
\tablewidth{0pc}
\tablecaption{Observation Log\label{tab:obslog}}
\tablehead{\colhead{Mission} & \colhead{ObsID} & \colhead{Startdate} & \colhead{Exposure (ks)\tablenotemark{a}} & \colhead{78\,d phase}}
\startdata
\xmm & 0145190201 & 2003-02-20 & \multirow{2}{*}{29 / 43}  & \multirow{2}{*}{0.15--0.26}\\
\xmm & 0145190101 & 2003-02-28 & & \\\hline
\xmm & 0673920201 & 2012-02-05 & \multirow{2}{*} {19 / 32} & \multirow{2}{*} {0.03--0.08} \\
\xmm & 0673920301 & 2012-02-09 & & \\
\chandra & 12987 &  2012-02-11 & \multirow{2}{*} {29} & \multirow{2}{*} {0.11-0.12} \\
\chandra & 14391 & 2012-02-11 & & \\\hline
\xmm & 0724810401 & 2013-11-12 & 23 / 32 & 0.30--0.31 \\
\nustar & 30002039005 & 2013-11-12 &  124 & 0.30--0.33 \\\hline 
\xmm & 0729561301 & 2014-07-09 & 38 /  43 & 0.367--0.374 \\
\nustar & 80001042002 & 2014-07-09 & \multirow{2}{*} {132} &  \multirow{2}{*} {0.36--0.42} \\
\nustar & 80001042004 & 2014-07-12 & &
\enddata
\tablenotetext{a}{\xmm exposure time given for pn / MOS}
\end{deluxetable}%

\subsection{\nustar}
\label{susec:nustar}

\nustar \citep{harrison13a} data were extracted using the standard \texttt{nustardas} pipeline v1.6.0 as distributed with HEASOFT V6.19. We applied the CALDB 20160706. Source data were extracted from 50\asec regions centered on the source and within the pointing uncertainty for the J2000 coordinates of \ngc. Background spectra were extracted from annuli centered on the same coordinates with  inner radii of 90\asec and  outer radii of 200\asec. The source becomes background dominated around 10\,keV and 20\,keV for the 2013 and 2014 data, respectively. We therefore carefully checked that our choice of background region does not influence the results, finding consistent results when using a circular background region with a radius of 100\asec located elsewhere on the same chip. Note that we only use the second observation in 2013, as the source was not detected in the first one (W15).

We additionally extracted data taken in SCIENCE\_SC (mode 06), during which the optical bench star tracker does not provide a solution for aspect reconstruction. While the source is therefore not reconstructed accurately on the sky, the responses can still be calculated correctly. We used regions of the same size centered on the centroid coordinates of the visible point source in these datasets. See \citet{gx339IHS} and \citet{walton16a} for details on mode 06 extraction. Using these data we add about 10\% and 15\% to the 2013 and 2014 data, respectively. This extra exposure time is included in Table~\ref{tab:obslog}.

The 2014 data were obtained during two separate observations, about 3\,d apart (Table~\ref{tab:obslog}).  We initially analyzed each observation separately but did not find any significant differences between them. We therefore added both observations and treat them as one epoch for the remainder of this paper.

\subsection{\xmm}
\label{susec:xmm}

The \xmm \citep{xmmref} observations were reduced with the \xmm Science Analysis System (SAS) v14.0.0
following standard procedures. The raw data files were filtered using
\texttt{epchain} and \texttt{emchain} to produce cleaned event lists for
each of the EPIC-pn and EPIC-MOS detectors, respectively \citep{pnref,
mosref}. As recommended, we use only single and double events for EPIC-pn,
and single to quadruple events for MOS. We exclude periods of high
background flaring. Science products were then produced using  \texttt{xmmselect},
with the source emission extracted from circular regions of radius
$\sim$20--30\asec (depending on the source brightness and its proximity to
bad detector columns) and the background estimated from larger areas on the
same CCD free of other contaminating point sources. Redistribution matrices
and auxiliary response files were generated with \texttt{rmfgen} and \texttt{arfgen},
respectively. After performing the data reduction separately for each of
the two MOS detectors, we combined the data from these detectors into
a single spectrum for each epoch using the FTOOL \texttt{addascaspec}.

\subsection{\chandra}
\label{susec:chandra}

\chandra \citep{chandraref} observed \ngc for two back-to-back observations in 2012 (Table~\ref{tab:obslog}). We extracted the ACIS-S with the standard CIAO v4.8 pipeline. The source spectra were extracted from circular regions with 3\asec radius centered on the J2000 coordinates, the background from 25\asec radius regions to the north-east from a source-free area. The spectra of the two observations were added with the CIAO tool \texttt{combine\_spectra}. In addition, the \chandra observations were performed within a few days of the 2012 \xmm observations and did not show significant changes in spectral shape or flux. We therefore treat all the 2012 data as a single epoch in the following analysis.

\section{Spectral Analysis}
\label{sec:spec}

For all our spectral fits we assume a Galactic absorption column of $1.38 \times 10^{20}$\,cm$^{-2}$ \citep{kalberla05a}. We model the absorption with an updated version of the \texttt{tbabs} model \citep{wilms00a}, using the corresponding abundances and cross-sections from \citet{verner96a}. We allow for an additional absorption column intrinsic to the ULX or NGC\,5907, as found by \citet{sutton13a} and W15.

The data were fitted using the Interactive Spectral Interpretation System  \citep[ISIS v1.6.2,][]{houck00a}. We rebinned the \nustar data in ISIS to a signal-to-noise ratio (\snr) of 4 below 10\,keV and 3 above. We additionally rebinned the data by at least a factor of 3 to prevent oversampling of the energy resolution. The \xmm EPIC-pn data were rebinned to a \snr of 6 below 5\,keV and 4 above, while the MOS data were rebinned to a \snr of 5 below 5\,keV and 3 above. We use the \xmm data in the energy range between 0.5--10\,keV and \nustar between 3--40\,keV. The \chandra data were rebinned to a S/N of 4 between 0.5--10 keV.

\subsection{The 2014 epoch}

\begin{figure}
\begin{center}
\includegraphics[width=0.95\columnwidth]{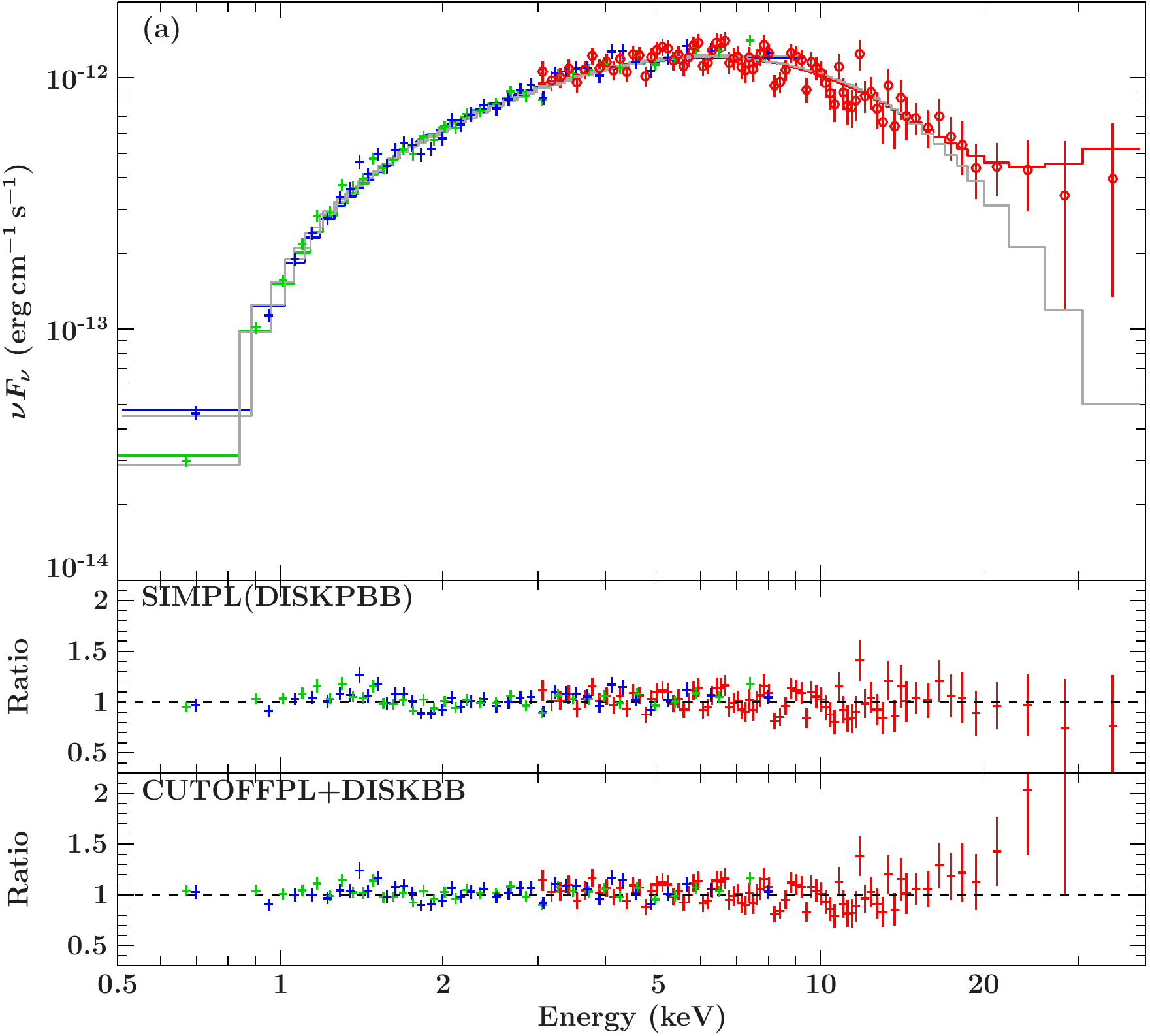}
\caption{Unfolded broadband spectrum of \ngc. \xmm EPIC-pn data are shown in green, MOS data are shown in blue, and the combined \nustar FPMA and FPMB data are shown in red. The best-fit \texttt{SIMPL$\times$diskpbb} model is superimposed. The gray line shows the  \texttt{cutoffpl+diskbb} model.  Residuals for the different models in terms of data-to-model ratio are shown in the lower panels. The corresponding parameter values are given in Table~\ref{tab:specres}. Data were rebinned for visual clarity. }
\label{fig:obs3spec}
\end{center}
\end{figure}

The 2014 data clearly show a hard spectrum with a visible turnover at high energies (Figure~\ref{fig:obs3spec}). Neither a pure power power-law ($\chi^2=1367/667$\,dof) nor  a simple multi-color blackbody of a geometrically thin, optically thick disk accretion disk (\texttt{diskbb}, $\chi^2=997/667$\,dof)  adequately describe this shape. This is very similar to other ULXs studied by \nustar and to previous studies of \ngc \citep{sutton13a, israel16b}. 

A phenomenological  cutoff  power-law model (\texttt{cutoffpl}), however, provides an acceptable fit with  $\chi^2=739$ for 666\,dof. 
This model can be improved by adding a \texttt{diskbb} model with a temperature of $\sim0.3$\,keV ($\chi^2=723$ for 664 d.o.f., $\Delta\chi^2=16$ for two additional parameters). Such a model is often used to describe Galactic binaries \citep[e.g.][]{mcclintock06a}.
However, while the \ngc spectrum is dominated by the power-law component, similar to the low/hard state of black hole binaries, the values we obtain are very different, e.g., the photon-index is much harder ($\Gamma=0.83^{+0.13}_{-0.15}$ instead of 1.4--1.8) and the folding energy much lower ($E_\text{fold}=5.3^{+0.7}_{-0.6}$\,keV instead of $\gg20$\,keV; see, e.g., \citealt{gx339}).

A more physically motivated model is the \texttt{diskpbb} model  \citep{mineshige94a}, which allows for a variation of the temperature gradient $p$ of the multicolor black-body, with $T(r)\propto r^{-p}$. Recent \nustar results have shown that ULX spectra are often well described with a temperature gradient somewhat shallower than the canonical $p=0.75$ expected for a thin disk \citep[e.g.,][]{bachetti13a, brightman16b}. Such shallower gradients are expected in sources with very high to super-Eddington luminosities, in which the accretion disk can increase its geometrical thickness due to radiation pressure and advection \citep{abramowicz88a}. We obtain a comparable fit to the cutoff power-law model with $\chi^2=740$ for 666 dof. We find $p=0.598^{+0.011}_{-0.010}$ and $T_\text{in}=3.46^{+0.17}_{-0.15}$\,keV.

However, this model leaves significant residuals at the highest energies. 
Such a hard energy excess is expected if a significant fraction of the thermal photons are Compton scattered to higher energies, resulting in an additional high-energy power law continuum. Evidence for an additional high-energy power law continuum has now been observed in several ULXs \citep[e.g.,][]{walton13a, walton14a, walton15b, mukherjee15a}.  Here we model the hard excess with the \texttt{simpl} model \citep{steiner09a}, which emulates up-scattering of the thermal seed photons into a power law tail. We find an excellent fit with $\chi^2=725$ for 664\,dof ($\Delta\chi^2=15$ for two additional parameters, Table~\ref{tab:specres}). We find a scattering fraction (i.e., flux in the power law tail) of $F_\text{sctr}\approx0.09$; however, this value is highly degenerate with the photon index. The corresponding $\chi^2$-landscape is complex so that a simple uncertainty estimation is not possible. By using the XSPEC \texttt{steppar} command in the 2D-space between $F_\text{sctr}$ and $\Gamma$, we find that $F_\text{sctr}$ has a lower limit of 0.036 at 90\% confidence. We do not find an upper limit due to the fact that $\Gamma$ can become very high, i.e., steep.

In all these models  a small excess around 1.5\,keV is visible, which can be linked to larger calibration uncertainties around the known ``silicon bump'' \citep{read14a}.

\begin{deluxetable*}{rlllll}
\tablewidth{0pc}
\tablecaption{Best-fit model parameters for all epochs.\label{tab:specres}}
\tablehead{\colhead{} & \multicolumn{2}{c}{2014} & \colhead{2013} & \colhead{2012} & \colhead{2003} \\
\colhead{Parameter}  & \colhead{Cpl+Diskbb} & \colhead{Simpl(Diskpbb)} & \colhead{Simpl(Diskpbb)} & \colhead{Diskpbb} & \colhead{Diskpbb}}
\startdata
 $ N_\text{H}~(10^{22}\,\text{cm}^{-2})$ & $0.85^{+0.14}_{-0.12}$ & $0.70\pm0.04$ & $0.65\pm0.08$ & $0.57\pm0.10$ & $0.76\pm0.06$ \\
 $ \mathcal{F}~(10^{-12}\,\text{erg}\,\text{cm}^{-2}\,\text{s}^{-1})\tablenotemark{a}$ & $2.54^{+0.26}_{-0.17}$ & $2.44^{+0.08}_{-0.32}$ & $0.86^{+0.09}_{-0.24}$ & $0.83\pm0.05$ & $1.93\pm0.08$ \\
 $ A_\text{disk}\tablenotemark{b}$ & $3.0^{+7.1}_{-2.3}$ & $\left(8.9^{+3.8}_{-2.8}\right)\times10^{-4}$ & $\left(2.8^{+6.5}_{-1.6}\right)\times10^{-4}$ & $\left(2.6^{+4.5}_{-2.0}\right)\times10^{-4}$ & $\left(4.9^{+4.6}_{-2.9}\right)\times10^{-4}$ \\
 $ \Gamma$ & $0.83^{+0.13}_{-0.15}$ & $1.1^{+1.7}_{-0.0}$ & $1.1^{+1.9}_{-0.0}$ & --- & --- \\
 $ E_\text{fold}~(\text{keV})$ & $5.3^{+0.7}_{-0.6}$ & --- & --- & --- & --- \\
 $ T_\text{in}~(\text{keV})$ & $0.30^{+0.09}_{-0.06}$ & $2.94^{+0.23}_{-0.30}$ & $2.7^{+0.5}_{-0.9}$ & $3.5^{+1.3}_{-0.7}$ & $3.1^{+0.7}_{-0.4}$ \\
 $ p$ & --- & $0.610^{+0.015}_{-0.013}$ & $0.552^{+0.028}_{-0.020}$ & $0.68^{+0.07}_{-0.05}$ & $0.585^{+0.022}_{-0.020}$ \\
 $ F_\text{sctr}$ & --- & $>0.036$ & $<0.11$\tablenotemark{c} & --- & --- \\
$\mathcal{L}_{14}~(10^{40}$\,erg\,s$^{-1}$)\tablenotemark{d} & $5.5^{+0.6}_{-0.4}$ & $5.25^{+0.18}_{-0.69}$ & $1.85^{+0.18}_{-0.51}$ & $1.79\pm0.11$ & $4.15^{+0.17}_{-0.16}$ \\
$\mathcal{L}_{17}~(10^{40}$\,erg\,s$^{-1}$)\tablenotemark{e} & $8.8^{+0.9}_{-0.6}$ & $8.51^{+0.28}_{-1.12}$ & $3.00^{+0.30}_{-0.82}$ & $2.90^{+0.18}_{-0.17}$ & $6.72^{+0.28}_{-0.26}$ \\
\hline$\chi^2/\text{d.o.f.}$   & 723.52/664& 725.49/664& 268.09/231& 197.78/214& 382.06/411\\$\chi^2_\text{red}$   & 1.090& 1.093& 1.161& 0.924& 0.930\enddata
\tablenotetext{a}{flux between 0.5--10\,keV}
\tablenotetext{b}{normalization of disk model in units of $\left(R_\text{in, km}/d_{10}\right)^2\cos(\theta)$}
\tablenotetext{c}{for a fixed value of $\Gamma=1.1$}
\tablenotetext{d}{luminosity between 0.5--10\,keV for a distance of 13.4\,Mpc}
\tablenotetext{e}{luminosity between 0.5--10\,keV for a distance of 17.1\,Mpc}
\end{deluxetable*}

\subsection{Comparison to archival observations}

To put the 2014 data into context with previous observations of \ngc, we perform our own analysis of the 2003, 2012,  and 2013 epochs (Figure~\ref{fig:xrtlc}, Table~\ref{tab:obslog}).
We fit the 2013 broad-band data with the same \texttt{simpl$\times$diskpbb} model used for the 2014 data. For the \xmm-only observations in 2003 and the \xmm plus \chandra observation in 2012 we did not include the \texttt{simpl} model, as the lack of coverage at high energies does not allow us to constrain its parameters. As can be seen in Figure~\ref{fig:obs3spec}, the hard tail modeled by the \texttt{simpl} model only becomes relevant above $\sim$15\,keV.
We also did not use the \texttt{diskbb + cutoffpl} model, as the power-law parameters could only be very weakly constrained with the soft data alone.

The best-fit values for all epochs are given in Table~\ref{tab:specres}. The 2012 and 2013 observations have a very similar 0.5--10\,keV flux, around $8.4\times10^{-13}\,\ergcms$, while both the 2003 and 2014 observations have a higher flux ($\sim2.5\times10^{-12}\,\ergcms$). However, neither the disk temperature nor the temperature gradient $p$ show a clear correlation with flux. Instead, the 2012 data  show a significantly higher value of  $p$ than the other observations, which  are all consistent with each other. With respect to the temperature, all observations are consistent within their uncertainties.

\begin{figure}
\begin{center}
\includegraphics[width=0.95\columnwidth]{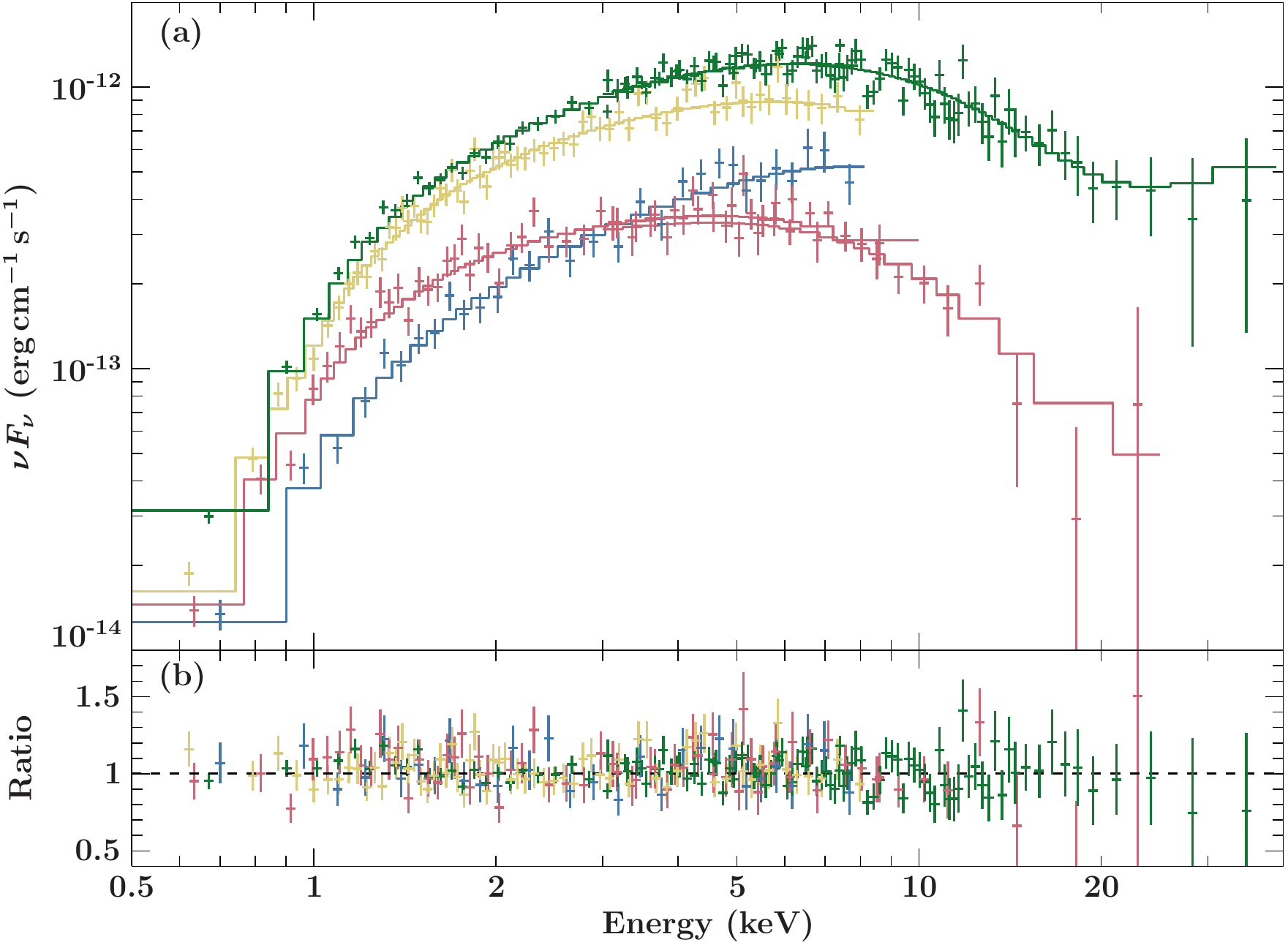}
\caption{Unfolded broadband spectrum of \ngc for all four epochs.  For clarity, we only show the \xmm EPIC-pn data and the \nustar FPMA data (where available). The 2003 data are shown in orange, the 2012 data in blue, the 2013 data in red and the 2014 data in green, together with their respective best-fit models. The lower panel shows the residuals to each model in terms of data-to-model ratio.}
\label{fig:allunfold}
\end{center}
\end{figure}

The difference can also be seen in Figure~\ref{fig:allunfold}, where we show the unfolded spectra of all four epochs. While the observations in 2012 and 2013 show very similar fluxes, their spectral shapes are distinctly different, with the 2012 flux rising much more steeply with energy.
On the other hand, the 2003 and 2014 data agree very well with each other.

\begin{figure}
\begin{center}
\includegraphics[width=0.95\columnwidth]{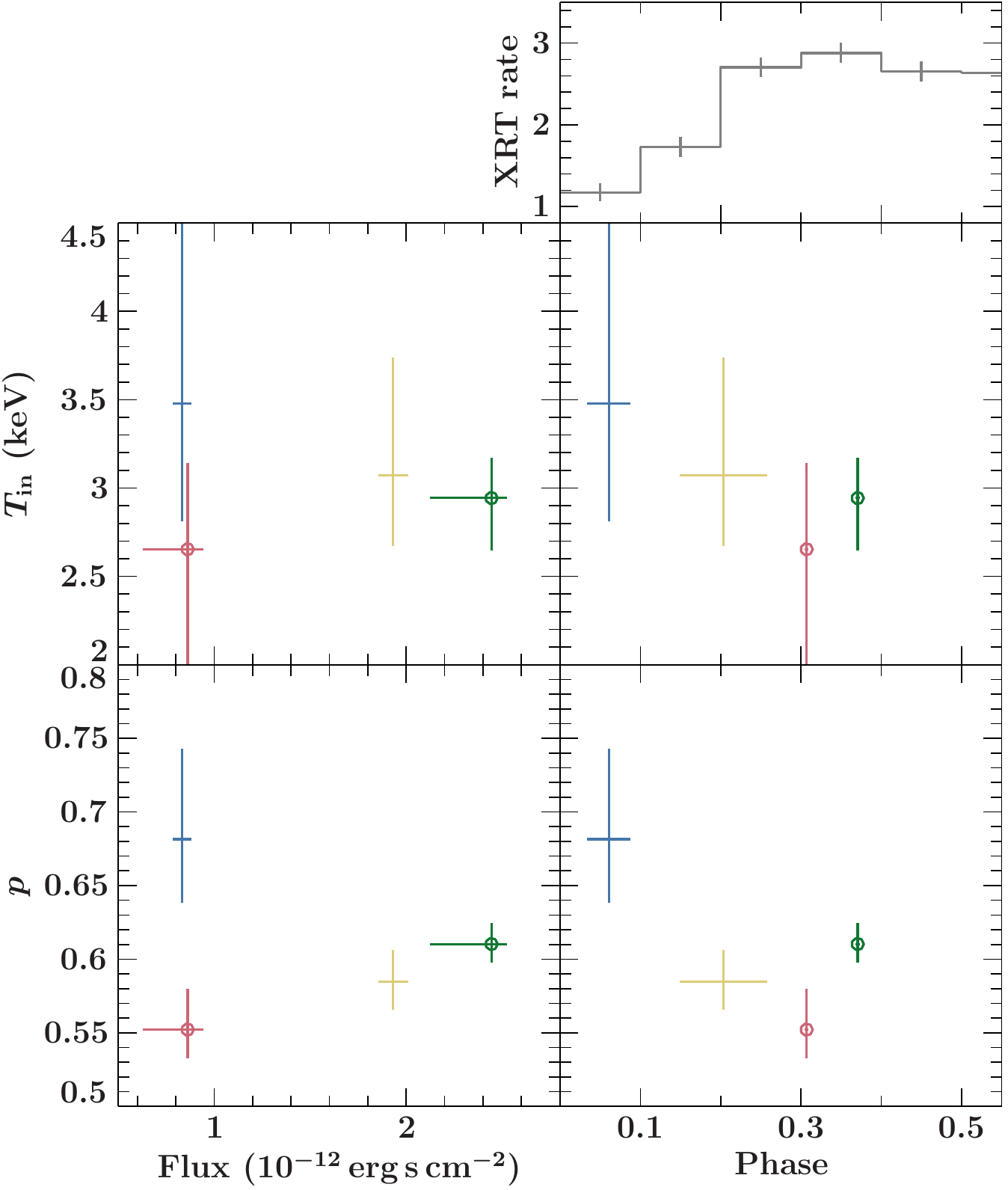}
\caption{Inner accretion disk temperature (top) and $p$ (bottom) for the four epochs as functions of 0.5--10\,keV flux (left) and 78\,d-phase (right). The observations taken with \nustar and \xmm are marked by circles and we use the same color-code for each observation as in Figures~\ref{fig:allunfold} and \ref{fig:cm}. The top of the right column shows the average \swift/XRT profile of the 78\,d-period in units of $10^{-2}$\,cts\,s$^{-1}$ (W16). 
}
\label{fig:parevol}
\end{center}
\end{figure}

We find a tentative trend of the spectral parameters as a function of 78\,d-phase. We list the phases of each epoch in Table~\ref{tab:obslog} and plot the values of $T_\text{in}$ and $p$ versus flux and phase, respectively, in Figure~\ref{fig:parevol}. The 2003, 2013, and 2014 data were all taken around phases 0.2--0.4 and show low values of $p$. On the other hand, the 2012 data were taken close to phase 0 and $p$ is significantly higher, albeit with large uncertainties. 

We note, however, that the phase of the 2003 data is relatively uncertain, as the period was extrapolated back by over 10 years. With an  estimated uncertainty of $\pm0.5$\,d on the period (W16),  the 2003 data can be located at phases between $-0.18$ and 0.58. For the following discussion we will nonetheless assume the predicted value of $\sim0.2$.

To check if the spectral variation with 78\,d-phase is stable over longer periods of time, we extracted spectra from the \swift/XRT monitoring campaign between 2014 March and 2016 July using the online \swift/XRT data products generator \citep{evans09a}. We selected data for the low-phase spectra between phases $0.7 \leq  \phi \leq 0.15$ and for the high phase between $0.3\leq\phi\leq0.65$, based on a period of $P=78.12$\,d and $T_0= 56663.0$\,MJD, the minimum of the profile. While the data quality of these XRT spectra are not sufficient to confirm the spectral changes seen in the \xmm data, they are fully consistent with the respective \xmm models. 

\begin{figure}
\begin{center}
\includegraphics[width=0.95\columnwidth]{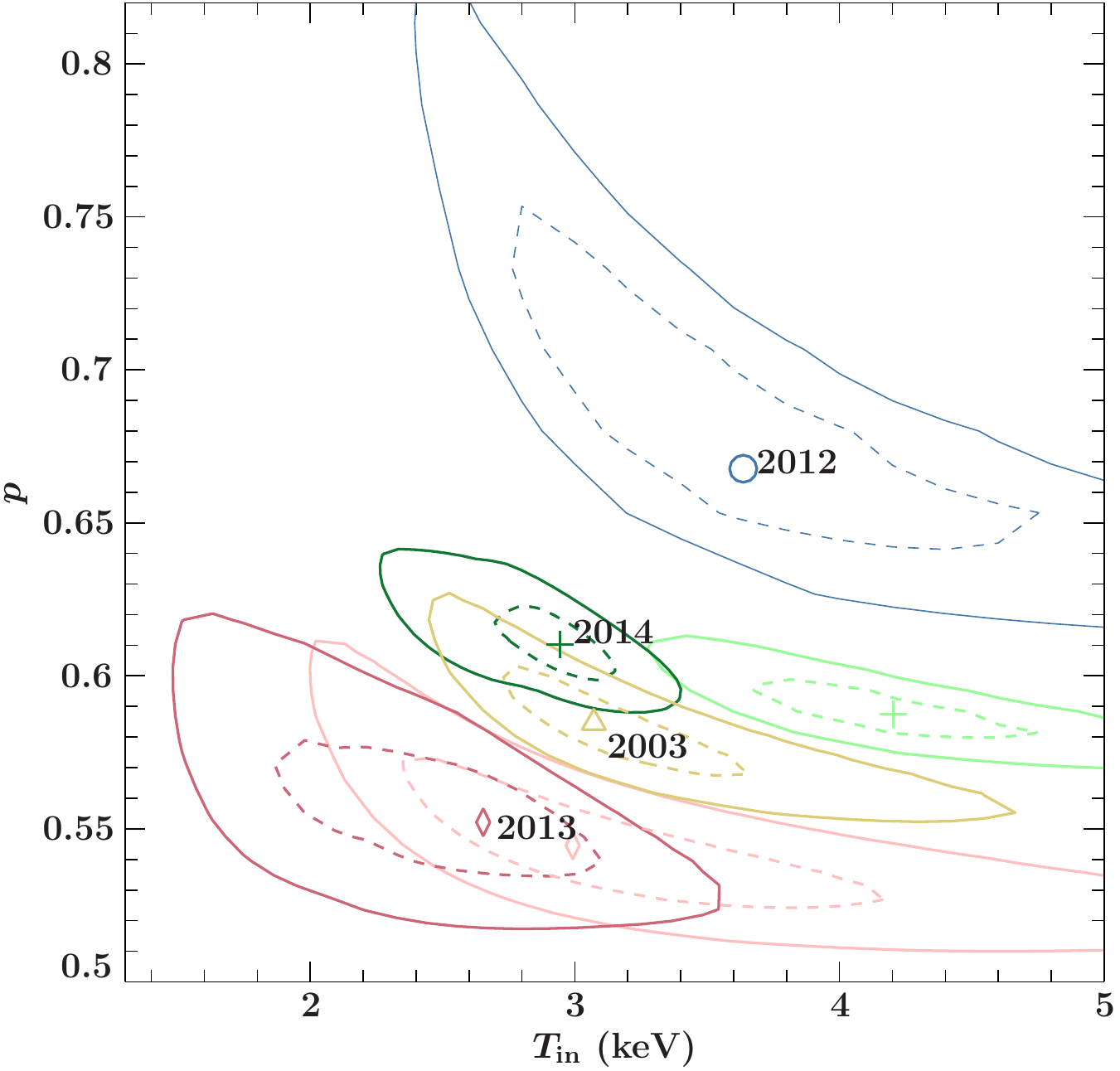}
\caption{Confidence contours in $\chi^2$-space for all four epochs between the inner accretion disk temperature $T_\text{in}$ and $p$. Solid lines indicate a $\Delta\chi^2$ of  9.21 (99\%),  and dashed lines indicate a $\Delta\chi^2$ of 2.30 (1-$\sigma$). The best-fit is marked by  a symbol. The color and symbols are as follows: 2003: orange, triangle; 2012: blue, circle, 2013: red, diamond, 2014: cross, green. The pink and light green lines show the results for  2013, and 2014 when only using \xmm, respectively.}
\label{fig:cm}
\end{center}
\end{figure}

To look for possible degeneracies between  $p$ and  temperature, we calculate confidence contours for these two parameters for each observation, shown in Figure~\ref{fig:cm}. We find that the 2003 and 2014 data are compatible with each other, while both the 2012 and 2013 data are significantly different. The 2012 data are consistent within their uncertainties with a standard temperature profile for a thin disk \citep[i.e., $p=0.75$,][]{shakura73a}. 

The uncertainties in $p$ are also influenced by the variability of the absorption column. If we fix the column at $6.7\times10^{21}$\,cm$^{-2}$, the average value of all four epochs, we find drastically smaller confidence contours while finding statistically comparable fits ($\Delta\chi^2 \le 3$ in all cases). All contours are fully within the contours presented in Figure~\ref{fig:cm}, however, we note that the 2012 data are no longer consistent with values $p\geq0.75$, thus ruling out a standard thin disk. As we do not know if the intrinsic absorption column has changed between observations, we continue our discussion with the model in which the column is  allowed to vary independently between observations.

We also show contours for 2013 and 2014 using the \xmm data only, to investigate if only using data below 10\,keV results in a systematic parameter shift. These contours are shown in  light color in Figure~\ref{fig:cm}. We find that the \xmm-only data seem to prefer a higher temperature, but find almost the same values for $p$. This is not surprising as the peak of the spectrum, which determines the temperature, is at the edge of the \xmm range, while $p$ is most relevant at lower energies and determines the slope of the spectrum up to the peak. Therefore the \xmm data drive $p$, while the temperature is mainly determined through the high-energy coverage of \nustar.

\section{Discussion and conclusions}
\label{sec:disc}

We have presented a detailed analysis of the broad-band spectrum of \ngc taken simultaneously by \nustar and \xmm during a bright phase in 2014 where the luminosity was $\sim8.5\times10^{40}$\,erg\,s$^{-1}$. We find that the spectrum is very well described by a multi-color black-body with a temperature profile with $p=0.608^{+0.014}_{-0.012}$, which is shallower than the expected value of $p=0.75$ for a thin disk \citep{shakura73a}. This suggests that advection plays an important role in the accretion disk during that observation.

At energies above $\sim$15\,keV we find evidence for a hard excess over the thermal continuum, which we describe by a power-law tail due to Comptonization of the disk photons. Even though the photon-index of this power-law could not be well constrained, this component provides a significant improvement in terms of $\chi^2$ ($\Delta\chi^2=15$ for two additional parameters). Given that the pulsed fraction is increasing with energy \citep{israel16b}, this power-law tail is likely connected to a hard continuum from the accretion column.

It is currently not known how the accretion disk is structured in neutron star ULXs. While for the often implied strong magnetic fields of these systems \citep[e.g.,][]{dallosso15a, mushtukov15a, tong15a}, a truncation of the disk at  $\sim$1400\,km radii is expected, due to the very high luminosity of \ngc the spherization radius is still about an order of magnitude larger than the magnetospheric radius \citep{king16a}. 
This implies that a geometrically thick disk is present, in which advection becomes important, in agreement with our spectral models. However, to sustain super-Eddington accretion rates, \citet{israel16b} argue that the disk has to be thin at the magnetospheric radius to not engulf the neutron star up to high magnetic latitudes. This problem could be mitigated by strong geometrical collimation, allowing the radiation to escape along a narrow funnel, which, however, is at odds with the observed smooth sinusoidal pulse profile \citep[see also][]{p13,bachetti14a}.

We put the 2014 data into context with observations taken in 2003, 2012, and 2013.  We find that in all observations the modeled disk temperature is similar, with $T_\text{in}\approx 2.75$\,keV. 
Assuming a slim-disk accretion geometry, we might therefore assume that we always see the hottest, innermost part of the accretion disk which is likely located close to the magnetospheric radius $r_\text{m}$. \citet{israel16b} argue that the surface magnetic field of \ngc should be around $3\times10^{13}$\,G, which implies $r_\text{m} \approx 1400$\,km \citep{cui97a}. If we assume that the observed radiation is dominated by the accretion disk, we can estimate the viewing angle from the normalization of the \texttt{diskpbb} model. 

We follow the description of \citet{soria15a} and \citet{brightman16b}. We assume a color correction factor of $\kappa=3$ and a geometric factor $\xi=0.353$, appropriate for the high Eddington fraction of \ngc \citep{soria15a}. We then calculate the inclination $\theta$ as:
 
\begin{equation}
\cos(\theta) = \frac{\xi^2\kappa^4N{d_{10}}^2}{{r_\text{m}}^2}
 \end{equation}
 
 Where $N$ is the normalization of the \texttt{diskpbb} model and $d_{10}$ is the distance in units of 10\,kpc.
 This results in an almost edge-on view of $\sim89\deg$.

This inclination is very high and makes it difficult to explain how we would be able to see the regions close to the compact object at all, as in super-Eddington accretion the accretion disk is expected to have some geometrical thickness, blocking our line-of-sight. The result depends on our assumptions of $\kappa$ and $\xi$, but we obtain high inclinations for all realistic assumptions \citep[e.g., $1.7 \leq \kappa \leq 3$,][]{watari03a}. About 20\% of the observed flux is pulsed \citep{israel16b}, indicating that it is produced by the rotating accretion column of the neutron star.  A lower flux from the disk, however, increases the estimated inclination further. 

Another large systematic uncertainty in this estimate is the strength of the magnetic field, and the location of the hottest part of the disk. \citet{king16a} argue that a low magnetic field ($\sim$$10^{10}$\,G) can explain the observed timing properties of M82~X-2, which are very similar to \ngc. With a lower magnetic field and a consequently much smaller magnetospheric radius the implied inclination would be smaller. In fact, in a face-on geometry ($\cos(\theta)=1$) we would imply a magnetic field of $B\approx 6\times10^{10}$\,G for the inner disk radius to be at the magnetospheric radius during the bright 2014 observation.

\subsection{The 2013 observation}

The 2013 observation shows a distinctly different spectrum than the 2012 observation, despite having a very similar flux. It also shows a significantly different spectrum than both the 2003 and 2014 observations, despite being located in phase between them. This behavior might be related to the fact that the 2013 data were obtained only  4\,d after the source's luminosity was below the detection limit of \xmm and \nustar (W15). 
This remarkable drop in luminosity is clearly not related to the stable 78\,d-period, and so far has not been seen to repeat (W16).
It is reasonable to expect that it is caused by the so-called ``propeller effect'' or the centrifugal inhibition of accretion, where the magnetospheric radius becomes larger than the corotation radius. This regime can be entered if the ram pressure of the accreting material drops,  leading to a further dramatic reduction in accretion rate and consequently luminosity. 
During such a state, the inner accretion disk would get depleted, and the slightly lower temperature measured could be an indication that the inner accretion disk 
was still in the process of refilling during the second 2013 observation in which the source was detected.

\subsection{Connection between the super-orbital period and spectral changes}
W16 argued that the 78\,d-period is most likely either orbital or super-orbital in nature. The new results by \citet{israel16b} indicate an orbital period of $\sim$5\,d, confirming the super-orbital nature of the 78\,d period.

The physical origin of super-orbital periods is often linked  to precession of the accretion disk. Such precession could explain the observed regular flux variations without changes in the physical conditions of the accretion flow, as our viewing angle would change periodically \citep[as also postulated for the ULX M82~X-2,][]{kong16a}. Using the ratio of the \texttt{diskpbb} normalizations and the geometric effect that the flux is reduced by $\sqrt{\cos(\theta)}$, under the assumption that the disk is relatively flat, we can calculate the required change in inclination angle $\theta$.  The largest variance in $\theta$ is required if we observe \ngc face-on ($\theta=0$) during the bright phase.
This would require an inclination angle of $\theta\approx 25^\circ$ during the faint phases (based on the 2012 data) and a half opening angle of the precession of $\sim13^\circ$. This is lower than the precession seen in SS\,433 \citep{khabibulllin16a}, which is often argued to be a Galactic example of super-Eddington accretion analogous to ULXs, but viewed close to edge-on such that the X-ray emitting regions are obscured from view \citep{fabrika04a}.
While a face-on view is in contradiction to the estimated viewing angle from the model normalization assuming a strong magnetic field, it shows that a varying viewing angle is a possible explanation of the observed flux changes for all inclinations.

While a precessing disk can naturally explain the differences in observed flux across the 78-day cycle, we need to understand how different spectral shapes can be measured at different viewing angles. From Figure~\ref{fig:parevol} it is clear that the strongest spectral change is observed in $p$.
At super-Eddington accretion rates it is expected that the accretion disk is flared up due to radiation pressure, i.e., becomes geometrically thick and increases in size with radius.
 In addition a strong wind is launched, which is also largely optically thick \citep{poutanen07a, dotan11a} and for which observational evidence has recently been found in NGC\,1313~X-1 and NGC\,5408~X-1\citep{pinto16a, walton16c}. The observed temperature profile therefore depends on which parts and with what angle we observe the accretion disk. Qualitatively we can envision a geometry where the apparent temperature gradient in the disk is changing as a function of viewing angle. Detailed calculations of this model are, however, beyond the scope of this paper.

\citet{israel16b} did not find pulsations in the 2012 and 2013 \xmm observations and give an upper limit of  a pulsed fraction of 12\%. 
Both these observations were taken at low apparent luminosities. If these changes are connected to real changes in accretion rate, the properties of the accretion column might change. For a lower accretion rate, as observed in 2013, the emission pattern of the accretion column might be wider, resulting in a reduced pulsed fraction.  

We have argued that the low observed flux of the 2012 data is not due to a lower intrinsic flux, but due to a change in viewing angle. This seems at first difficult to reconcile with the disappearing of  pulsations. However, it is also possible that the neutron star shows free precession, in step with the precession of the accretion disk \citep[e.g., as discussed for Her~X-1,][]{staubert13a}. In this case, the rotational axis at early super-orbital phases might be aligned close to our line of sight or, if the emission is axial-symmetric, close to 90\deg, also reducing the observable pulsed fraction.

It is clear that to understand super-Eddington neutron stars like \ngc more observational and theoretical work needs to be done.
For example, the currently available coverage of the phase-space is concentrated between phases 0--0.5. For a detailed test of the proposed spectral evolution and model, broad-band observations at later phases are necessary, which can be obtained with \xmm and \nustar. \ngc is an ideal target for these studies, given the stable and strong 78\,d-period. Its spectral similarities to other ULXs will help us understand this class of objects better and investigate how the type of compact object influences their behavior.

\acknowledgments
Based on observations obtained with \xmm, an ESA science mission with instruments and contributions directly funded by ESA Member States and NASA.
This work was supported under NASA Contract No. NNG08FD60C, and
made use of data from the {\it NuSTAR} mission, a project led by
the California Institute of Technology, managed by the Jet Propulsion
Laboratory, and funded by the National Aeronautics and Space
Administration. We thank the {\it NuSTAR} Operations, Software and
Calibration teams for support with the execution and analysis of these observations. This research has made use of the {\it NuSTAR}
Data Analysis Software (NuSTARDAS) jointly developed by the ASI
Science Data Center (ASDC, Italy) and the California Institute of
Technology (USA). 
This work made use of data supplied by the UK Swift Science Data Centre at the University of Leicester.
This research has made use of a collection of ISIS functions (ISISscripts) provided by ECAP/Remeis observatory and MIT (\url{http://www.sternwarte.uni-erlangen.de/isis/}).
We would like to thank John E. Davis for the \texttt{slxfig} module, which was used to produce all figures in this work.
The \swift/BAT transient monitor results were
provided by the \swift/BAT team.

\textit{Facilities:} \facility{NuSTAR}, \facility{XMM}, \facility{Chandra}, \facility{Swift}


\end{document}